\documentclass[11pt]{article}
\usepackage[dvips]{graphics}
\usepackage{natbib}  %To facilitate proper referencing a la Mueller et
\bibpunct{(}{)}{;}{a}{}{;}  %To change format of citation - no comma
\def\simless{\mathbin{\lower 3pt\hbox
   {$\rlap{\raise 5pt\hbox{$\char'074$}}\mathchar"7218$}}}   % < or of order
\def\simgreat{\mathbin{\lower 3pt\hbox
   {$\rlap{\raise 5pt\hbox{$\char'076$}}\mathchar"7218$}}}   % > or of order

\begin{document}

\title{\bf{Formation of dSph Satellites without Dark Matter}}
\author{Pavel Kroupa\thanks{e-mail: pavel@ita.uni-heidelberg.de; to
appear in Dynamics of Galaxies and Galactic Nuclei,
Proc. Ser. I.T.A. no. 2, eds. W. Duschl \& C. Einsel, Heidelberg}\\ 
Institut f\"ur Theoretische Astrophysik\\ Universit{\"a}t
Heidelberg\\ Tiergartenstr. 15, D--69121 Heidelberg\\ Germany}
\maketitle
%%%%%%%%%%%%%%%%%%%%%%%%%%%%%%%%%%%%%%%%%%%%%%%%%%%%%%%%%
\begin{abstract}
Recently it was shown that low-mass galaxies containing no dark matter
can naturally evolve to dSph-like satellites in a tidal field. Such
dSph-like satellites fake total domination by dark matter. If this is
correct then it must be possible to assemble appropriate precursors of
dSph satellites. The genesis of dwarf galaxies in tidal tails appears
to be a likely possibility. This is documented by theoretical work and
observational surveys.  However, it is not yet clear if the Milky Way
could have sustained one or two mergers or fly-bys that were mild
enough not to have destroyed the Galactic disc. The tidal arms could
have been pulled from the incoming galaxy. Also, this scenario has to
account for the correlation between brightness and metal abundance
displayed by the Galactic dSph satellites, and lead to at least some
debris on bound eccentric orbits. 
\end{abstract}
%%%%%%%%%%%%%%%%%%%%%%%%%%%%%%%%%%%%%%%%%%%%%%%%%%%%%%%%%

\section{Introduction}
\label{sec:intro}
Galaxy interactions are common at the present epoch. Nearby examples
are the Magellanic-Clouds---Milky-Way (e.g. \citealt{deB97}) and the
Sagittarius-dwarf-galaxy---Milky-Way \citep{Ietal97} systems.  Galaxy
interactions were much more common in the past though, and the Milky
Way probably carries some scars from a more active past. Evidence for
this is the probable existence of the thick disc \citep{GWK89,GWJ95,
Retal96, Wetal96} as well as probable tidal debris in the halo
(e.g. \citealt{M94,LL95}). However, the number and strength of
collisions of the Milky Way with small companion galaxies remains to
be quantified. Interesting limits are provided by the metallicity
distribution of halo stars \citep{Uetal96} and the longevity of halo
phase-space structure \citep{JHB96}.

The suspected tidal debris surrounding the Milky Way consists of
individual globular clusters and nine known dwarf spheroidal (dSph)
satellite galaxies, which are approximately aligned on a few great
circles. Their masses ($\simless10^8\,M_\odot$, e.g. \citealt{IH}) are
too small for dynamical friction in an extended Galactic dark halo to
significantly affect their distribution over a Hubble time. Their
orbital distribution must be primordial, apart from orbital precession
in a possibly non-spherical dark halo (e.g. \citealt{FJ}).

Inferring the genesis of the dSph satellites is of greatest
importance. Their kinematical, structural and photometric properties
can be interpreted to imply that these small ($\simless1-2$~kpc) and
very faint ($\simless10^7\,L_\odot$) collision-less stellar systems
are completely dark-matter dominated, in which case they ought to be
relics from early cosmic structure formation. The alternative
interpretation is that the dSph satellites are remnants of galaxies
without dark matter, apart from burned-out stars \citep{K97,KK}. Such
remnants arise naturally after removal of roughly 99~per cent of the
stars through periodic tidal harassment, if their orbital
eccentricities are large ($\simgreat0.5$). The proper motion of two
dSph satellites has been measured recently \citep{SI,Setal,SC}.
However, present estimates of orbital eccentricities are,
unfortunately, too inaccurate to exclude the alternative
interpretation, because the combined uncertainties in proper-motion
measurements and in the mass distribution of the Milky Way are too
large (e.g. \citealt{KB97}).  For recent reviews on the dark matter
content of dSph galaxies see \citet{M98} and \citet{O98}.

Both interpretations have significant implications for the nature of
dark matter. The former would imply that large amounts of dark matter
($\approx10^7-10^8\,M_\odot$) must be containable within the tight
phase-space limits defined by the dSph satellites. The latter
interpretation is consistent with the absence of dark matter in
reasonably well-understood dynamical systems that are smaller in
extend than a few~kpc. Examples of such systems are the Galactic disc
(e.g. \citealt{K91}) and star clusters. Globular clusters contain
about the same number of stars as dSph satellites, but they cannot be
their precursors. This is because the metallicity distribution and the
star-formation history are irreconcilably more complex in dSph
satellites (e.g. \citealt{G97,DC}). Also, globular clusters with
$\simgreat10^6$ stars and half-light radii $\simless10$~pc are stable
against tidal destruction over a Hubble time because their tidal radii
are $\simgreat90$~pc, even if the perigalactic distance is $R_{\rm
peri}\approx5$~kpc \citep{OLA}.

The precursor of a dSph satellite must, if this latter interpretation
is correct, have (i) essentially no dark matter, (ii) $10^7-10^9$
stars for significant tidal de-population within a Hubble time to
leave a recognisable {\it remnant}, (iii) a characteristic radius
roughly comparable to the tidal radius at peri-galacticon for
susceptibility to tidal damage, and (iv) an apparently complex
star-formation and metal-enrichment history. Such a system is {\it
collision-less}, because the half-mass relaxation time is longer than
a Hubble time.

Can such a dwarf galaxy be created naturally? The hint provided by the
significantly non-isotropic distribution of the Galactic dSph
satellites may be the key to the answer \citep{L82}.  Theoretical
computations that shed light on the formation of satellite galaxies in
tidal tails are discussed in Section~\ref{sec:mods}. Observational
evidence is presented in Section~\ref{sec:obs}, and
Section~\ref{sec:concl} contains the conclusions.

\section{Models}
\label{sec:mods}
Tidal tails are pulled out from galaxies if these are tidally
disturbed.  The length and mass of the tidal tails depend on the
strength of the interaction and on the mass and extend of the dark
halos surrounding the interacting galaxies \citep{DHM}.  Dwarf
galaxies, with masses $\simgreat10^8\,M_\odot$, are expected to form
within tidal tails, because only objects with such masses can become
gravitationally unstable towards collapse during a sufficiently strong
galaxy interaction that churns up the stellar and gaseous velocity
dispersions \citep{EKT}. Also, the interaction of gas-rich tidal tails
with the inter-galactic medium may lead to gravitational instabilities
near the ends of tidal tails \citep{BH}. The tidal field of the merged
or merging galaxies also limits the mass of self-gravitating
tidal-tail clumps.

{\it If}, {\it how} and {\it where} tidal tails produce
self-gravitating clumps is an extremely hard question to study
theoretically because of computational limitations. The largest number
of particles that can be used to represent a disc galaxy is, at
present, roughly $10^5$.  Grid methods in principle allow more
particles, but are not so well suited for the simulation of non-planar
tidal deformations that can extend over distances larger by
orders-of-magnitude than the thickness of the unperturbed galaxy's
disc.  The particle density along a tidal tail is small, and it is
difficult to asses if the condensations that form within the computer
tails are physical or merely artifacts that result from local Poisson
fluctuations and swing amplification (see e.g. \citealt{BT}). Gas
plays a major role in the formation of self-gravitating objects in
tidal tails, because dissipation forces the gas to accumulate near the
centre of a gravitational instability thereby enhancing it and leading
to star-formation. However, the numerical treatment of hydrodynamics
is a serious challenge, and usually recourse to simple but
computationally tractable approximations is required.

\citet{BH} use a combined N-body and smooth-particle-hydrodynamics
(SPH) code to model self-consistently two colliding Milky-Way-type
disc galaxies. The galaxies are each represented by about 37000
stellar and 8000 SPH particles. The trajectory is initially parabolic
with a pericenter distance of 8~kpc, and the discs are not
co-planar. Extended tidal features develop after the first
encounter. The relative orbit decays due to dynamical friction between
the dark halos and the galaxies meet again about 380~Myr after the
first passage. Coalescence occurs approximately 500~Myr after the
first passage, but the tidal tails continue expanding. In the
expanding tidal tails, \citet{BH} count 23~bound condensations that
contain at least 12~particles ($\simgreat10^8\,M_\odot$). The objects
condense at distances larger than about 40~kpc from the merged
galaxies.  The most massive of these consists of about 360~particles,
has a half-mass radius of 760~pc and a mass of approximately
$4\times10^8\,M_\odot$, 25~per cent of which is in the form of gas
that is more centrally concentrated than the stellar particles. Less
than 5~per cent of the mass of this object is in the form of dark
matter from the parent galaxy's halo.

\citet{EKT} use a two-dimensional N-body and sticky-particle code. A
self-consistent particle and gas disc is embedded in a rigid bulge and
dark halo potential. About 63000 stellar particles and 31000 sticky
particles represent the disc.  The perturber is approximated by a
point mass with the same mass as the galaxy, and is allowed to pass by
in the disc plane on a hyperbolic orbit with a pericenter distance of
about 20~kpc. Tidal tails form quickly and wrap up in the inner parts
of the disc where the gas particles sink to the centre. In the outer
parts of the disc the tidal arms expand. The overall appearance is
strikingly similar to the Hubble Space Telescope images of the
Antennae galaxies \citep{WS95,Wetal98}.  Many self-gravitating objects
condense in the stretching tidal arms, the most massive of which
consists of about 250~gas particles with a mass of about
$10^8\,M_\odot$.  This object is composed mostly of gas, with old
stellar particles contributing only about 40~per cent. Further
simulations show that in order for tidal-tail condensations with
masses of about $10^7\,M_\odot$ to form, perturber masses of at least
about 20~per cent of the galaxy mass are needed. However, it may be
possible that less massive tidal-tail condensations develop for even
smaller perturbers in simulations with higher resolution.  The authors
estimate that the mass of the perturber should be about 1.4~times the
mass of the galaxy for the outer-most tidal-tail condensations to be
expelled into inter-galactic space.

That the tidal debris, including the freshly assembled dwarf galaxies,
can end on bound eccentric orbits about the merged remnant galaxy is
demonstrated by \citet{HM}.

\section{Observations}
\label{sec:obs}
That star-formation on dwarf-galaxy scales occurs in tidal tails has
been known for some time. For example, \citet{MDL} show that a
gas-rich dwarf galaxy appears to be forming about 100~kpc from the
parent galaxy at the tip of one tidal tail in the Antennae
galaxies. This dwarf galaxy has a mass of about $10^9\,M_\odot$, as
estimated from the HI gas content and from the velocity dispersion
using 21~cm observations, and a diameter of about 15~kpc, and is
actively forming OB~stars.  In an R-band CCD imaging survey of
42~Hickson compact groups, \citet{HCZ} find 42~candidate dwarf
galaxies in tidal tails. These dwarfs have masses of
$10^6-10^8\,M_\odot$ estimated by assuming a mass-to-light ratio of
unity, have diameters of a few kpc and lie at projected distances of a
few tens of kpc from the parent galaxies. The authors estimate that
such dwarf galaxies contribute significantly (30--50~per cent) to the
number of dwarf galaxies in groups. Further observational evidence for
the formation of giant ($\approx10^8\,M_\odot$) gas clouds in
interacting galaxies is discussed by \citet{EKT}.

In their very interesting review of observational evidence for the
formation of dwarf galaxies in tidal tails, \citet{DM} also consider
the metallicity--brightness correlation.  Bright Galactic dSph
satellites are more metal rich than faint satellites, and this
correlation is also evident for elliptical galaxies and isolated dwarf
galaxies \citep{GW94,FB94}. \citet{DM} show that their sample of tidal
dwarf galaxies does not exhibit this correlation. The tidal dwarfs are
relatively metal rich (approximately one third solar) because they
inherit the pre-enriched gas from the parent disc galaxy.

\section{Conclusions}
\label{sec:concl}
Interacting gas-rich disc galaxies expel tidal tails that may fragment
at distances of many tens of ~kpc from the parent galaxies. The
formation of such tidal-tail fragments is now well documented in
theory and by observation. The fragments are gas rich, and form
massive stars hundreds of~Myr after the interaction. Self-consistent
models with dark matter demonstrate that the dark matter content of
such fragments is insignificant, and the simulations also suggest that
the fragments are gravitationally bound objects. The masses of the
fragments lie in the range $10^7-10^9\,M_\odot$ and they have
diameters of 1--10~kpc. Galaxy interactions involving any mass ratio
can lead to the formation of fragmenting tidal tails. However, only
interactions involving two large galaxies, one of them by necessity a
disc galaxy, will form massive star-forming tidal-tail condensations
that may be expelled into inter-galactic space.

The tidal-tail dwarf galaxies consist of a complex mixture of stellar
populations. They are composed of stars born in the parent galaxy as
well as of stars formed during the formation of the tidal tail
condensation. Complex self-enrichment is likely to occur, and it is
perhaps possible that some episodic star formation over time-scales of
Gyrs may result if the tidal dwarf falls back towards the merger
remnant thereby acquiring additional gas. Such a bound tidal-tail
dwarf galaxy will be repeatedly harassed by tidal forces, and will
ultimately evolve to a dSph-like satellite. Small isolated spheroidal
galaxies, such as Tucana, would be essentially without dark matter if
they are tidal debris on unbound orbits. Such systems should have
small dynamical $M/L$ ratios.

It thus seems natural to interpret at least some of the Galactic dSph
satellites as dark-matter-less remnants of such tidal-tail dwarf
galaxies.

However, significant problems remain. It is unclear if the observed
tidal-tail ``dwarf galaxies'' are really gravitationally bound objects
once star-formation ceases. Projects are under way with R. Andersen
and M. Fellhauer to investigate if vigorous star formation and
assembly in a tidal field can produce such systems.  Furthermore, it
must be demonstrated that such aged dwarfs have sizes and masses that
make them possible precursors of Galactic dSph satellites, if the
latter are to have been born in an early interaction of the Milky Way
with some other galaxy. Also, it is not clear if the early Milky Way
could have suffered an interaction with another galaxy that was
sufficiently strong to produce tidal arms, but not destructive of the
Galactic disc. The early gas-rich Galactic disc may have reformed
rapidly after such a perturbation, and the central bulge may have
acquired mass from the radial gas flows during the interaction.
Furthermore, it must be shown that tidal debris remains bound to the
Milky Way on orbits with eccentricities $\simgreat0.5$.  Finally, it
is at present not clear how the observed correlation between
brightness and metal abundance of the dSph satellites can be accounted
for. Metallicity-gradients in the early Galactic disc and
self-enrichment, in dependence of the tidal-dwarf galaxy's mass, may
play a role, but the small masses make gas confinement difficult.

%%%%%%%%%%%%%%%%%%%%%%%%%%%%%%%%%%%%%%%%%%%%%%%%%%%%%%%

\end{document}